\documentclass[aps,pra,preprint,12pt]{revtex4}
\usepackage{amsfonts}
\usepackage{amsmath}

\input{tcilatex}

\begin{document}

\title{Squeezing spectra from s-ordered quasiprobability distributions.
Application to dispersive optical bistability}
\author{Ferran V. Garc\'{\i}a--Ferrer, Isabel P\'{e}rez--Arjona, Germ\'{a}n
J. de Valc\'{a}rcel, and Eugenio Rold\'{a}n}
\affiliation{Departament d'\`{O}ptica, Universitat de Val\`{e}ncia, Dr. Moliner 50,
46100--Burjassot, Spain}

\begin{abstract}
It is well known that the squeezing spectrum of the field exiting a
nonlinear cavity can be directly obtained from the fluctuation spectrum of
normally ordered products of creation and annihilation operators of the
cavity mode. In this article we show that the output field squeezing
spectrum can be derived also by combining the fluctuation spectra of any
pair of $s$--ordered products of creation and annihilation operators. The
interesting result is that the spectrum obtained in this way from the
linearized Langevin equations is exact, and this occurs in spite of the fact
that$\ $no $s$--ordered quasiprobability distribution verifies a true
Fokker--Planck equation, i.e., the Langevin equations used for deriving the
squeezing spectrum are not exact. The (linearized) intracavity squeezing
obtained from any $s$--ordered distribution is also exact. These results are
exemplified in the problem of dispersive optical bistability.
\end{abstract}

\maketitle

\section{Introduction}

An appropriate tool for studying the fluctuations of a quantum light field $%
\hat{E}\left( t\right) $ is their spectrum, which is defined as the Fourier
transform of the field correlations $\left\langle \hat{E}\left( t+\tau
\right) ,\hat{E}\left( t\right) \right\rangle $, where $\left\langle
U,V\right\rangle =\left\langle UV\right\rangle -\left\langle U\right\rangle
\left\langle V\right\rangle $. Here we are concerned with the study of the
amount of squeezing provided by nonlinear cavities, i.e., optical cavities
containing a nonlinear medium and pumped by some input field. Nonlinear
cavities are known to produce large amounts of squeezed light for a
particular frequency or band of frequencies in the field exiting the cavity,
as a result of the interference at the cavity output mirror between the
partially squeezed intracavity mode and the reservoir modes \cite%
{Loudon87,Meystre91,Walls94,Drummond04,Gardiner00}.

When calculating the fluctuations of the output field quadratures, one uses
the squeezing spectrum \cite{Collet84,Collet85} defined as%
\begin{equation}
S_{\varphi }^{\mathrm{out}}\left( \omega \right) \equiv \frac{1}{4}%
+\dint\limits_{-\infty }^{+\infty }d\tau e^{-i\omega \tau }\left\langle :%
\hat{X}_{\varphi }^{\mathrm{out}}\left( t+\tau \right) ,\hat{X}_{\varphi }^{%
\mathrm{out}}\left( t\right) :\right\rangle ~,  \label{Sout}
\end{equation}%
where the field quadrature%
\begin{equation}
\hat{X}_{\varphi }^{\mathrm{out}}\left( t\right) \equiv \frac{\hat{a}_{%
\mathrm{out}}\left( t\right) e^{-i\varphi }+\hat{a}_{\mathrm{out}}^{\dagger
}\left( t\right) e^{i\varphi }}{2},  \label{quadratures}
\end{equation}%
being $\hat{a}_{\mathrm{out}}^{\dagger }$ and $\hat{a}_{\mathrm{out}}$ the
creation and annihilation operators of the output field that verify $\left[ 
\hat{a}_{\mathrm{out}}\left( t\right) ,\hat{a}_{\mathrm{out}}^{\dagger
}\left( t^{\prime }\right) \right] =\delta \left( t-t^{\prime }\right) $,
and $\varphi $ an arbitrary phase. In Eq. (\ref{Sout}) the label "$\mathrm{%
out"}$ refers to the field exiting the nonlinear cavity, $:\,:$ denotes
normal and time ordering, and the term $\frac{1}{4}$ corresponds to the shot
noise level.

\subsection{Squeezing spectrum from the generalized $P$ distribution}

The use of Eq. (\ref{Sout}) requires relating the correlations of the field
outside the cavity --which are the ones that are actually detected-- with
those of the intracavity field, which are readily calculated by solving the
master equation for the particular system. This relation is given by the
input--output theory \cite{Gardiner00}, which, when the nonlinear cavity is
fed with a coherent or vacuum field, states that \cite{Collet84}%
\begin{align}
\left\langle :\hat{a}_{\mathrm{out}}\left( t+\tau \right) ,\hat{a}_{\mathrm{%
out}}\left( t\right) :\right\rangle & =\gamma _{\mathrm{out}}\left\langle :%
\hat{a}\left( t+\tau \right) ,\hat{a}\left( t\right) :\right\rangle ,
\label{in-out1} \\
\left\langle :\hat{a}_{\mathrm{out}}^{\dagger }\left( t+\tau \right) ,\hat{a}%
_{\mathrm{out}}\left( t\right) :\right\rangle & =\gamma _{\mathrm{out}%
}\left\langle :\hat{a}^{\dagger }\left( t+\tau \right) ,\hat{a}\left(
t\right) :\right\rangle ,  \label{in-out2}
\end{align}%
where $\gamma _{\mathrm{out}}$ represents the cavity loss rate of the field
intensity at the output mirror, and $\hat{a}^{\dagger }$ and $\hat{a}$ are
the creation and annihilation operators of the intracavity field, which
verify $\left[ \hat{a}\left( t\right) ,\hat{a}^{\dagger }\left( t\right) %
\right] =1$. We recall that no simple relations like Eq. (\ref{in-out1},\ref%
{in-out2}) exist that relate correlations between outgoing fields and
intracavity fields unless those are calculated in normal order.

The dynamics of the nonlinear cavity is usually described, although not
necessarily, through Langevin equations obtained from a Fokker--Planck
equation. Then the form of Eqs. (\ref{in-out1},\ref{in-out2}) suggests the
use of the generalized $P$ representation \cite{Drummond80}, which we denote
by $\mathcal{P}$, as in this case%
\begin{align}
\left\langle :\hat{a}\left( t+\tau \right) ,\hat{a}\left( t\right)
:\right\rangle & =\left\langle \alpha \left( t+\tau \right) ,\alpha \left(
t\right) \right\rangle _{\mathcal{P}}, \\
\left\langle :\hat{a}^{\dagger }\left( t+\tau \right) ,\hat{a}\left(
t\right) :\right\rangle & =\left\langle \beta \left( t+\tau \right) ,\alpha
\left( t\right) \right\rangle _{\mathcal{P}},
\end{align}%
where $\alpha $ and $\beta $ are independent \textit{c}-numbers associated
to $\hat{a}$ and $\hat{a}^{\dagger }$ respectively that verify $\left\langle
\beta \right\rangle _{\mathcal{P}}=\left\langle \alpha \right\rangle _{%
\mathcal{P}}^{\ast }$, and $\left\langle f\right\rangle _{\mathcal{P}}$
denotes the average value of any function $f\left( \alpha ,\beta \right) $
calculated in the $\mathcal{P}$ representation. Then, making use of the
above equations one can write the squeezing spectrum as%
\begin{align}
S_{\varphi }^{\mathrm{out}}\left( \omega \right) & =\frac{1}{4}+\gamma _{%
\mathrm{out}}\mathcal{V}_{\mathcal{P}}\left( \omega ,\varphi \right) ~,
\label{Sout3} \\
\mathcal{V}_{\mathcal{P}}\left( \omega ,\varphi \right) & \equiv
\dint\limits_{-\infty }^{+\infty }d\tau e^{-i\omega \tau }\left\langle
X_{\varphi }\left( t+\tau \right) ,X_{\varphi }\left( t\right) \right\rangle
_{\mathcal{P}},  \notag
\end{align}%
with $X_{\varphi }\left( t\right) \equiv \left[ \alpha \left( t\right)
e^{-i\varphi }+\beta \left( t\right) e^{i\varphi }\right] /2$, which is a
well known and widely used result.

\subsection{Squeezing spectrum from other quasiprobability distributions}

The above presentation suggests that the use of the $\mathcal{P}$
distribution is mandatory in order to derive the squeezing properties of the
output field. (The usual Glauber--Sudarshan $P$ distribution presents well
known problems that will be recalled below.) Nevertheless, and this is at
the heart of our work, a suitable combination of two $s$-ordered \cite%
{Cahill69} two--time correlations, for example antinormally ($s=-1$) and
symmetrically ($s=0$) ordered as obtained by using the $Q$ (Husimi) and $W$
(Wigner) representations \cite{Walls94,Gardiner00,Carmichael99}, leads to
the same result.

Consider the obvious property%
\begin{equation}
2\left\langle \alpha \left( t+\tau \right) ,\alpha \left( t\right)
\right\rangle _{W}=\left\langle \alpha \left( t+\tau \right) ,\alpha \left(
t\right) \right\rangle _{\mathcal{P}}+\left\langle \alpha \left( t+\tau
\right) ,\alpha \left( t\right) \right\rangle _{Q},  \label{two-timeW}
\end{equation}%
where the subscript indicates the quasiprobability distribution used for
obtaining the correlations. Then it follows that $\left\langle :\hat{a}%
\left( t+\tau \right) ,\hat{a}\left( t\right) :\right\rangle =2\left\langle
\alpha \left( t+\tau \right) ,\alpha \left( t\right) \right\rangle
_{W}-\left\langle \alpha \left( t+\tau \right) ,\alpha \left( t\right)
\right\rangle _{Q}$, and then $\left\langle :\hat{X}_{\varphi }^{\mathrm{out}%
}\left( t+\tau \right) ,\hat{X}_{\varphi }^{\mathrm{out}}\left( t\right)
:\right\rangle =\gamma _{\mathrm{out}}\left[ 2\left\langle X_{\varphi
}\left( t+\tau \right) ,X_{\varphi }\left( t\right) \right\rangle
_{W}-\left\langle X_{\varphi }\left( t+\tau \right) ,X_{\varphi }\left(
t\right) \right\rangle _{Q}\right] $, so that the squeezing spectrum (\ref%
{Sout}) can be written as%
\begin{equation}
S_{\varphi }^{\mathrm{out}}\left( \omega \right) =\frac{1}{4}+\gamma _{%
\mathrm{out}}\left[ 2\mathcal{V}_{W}\left( \omega ,\varphi \right) -\mathcal{%
V}_{Q}\left( \omega ,\varphi \right) \right] ,  \label{Sout4}
\end{equation}%
where the notation is self--explicative, see Eq. (\ref{Sout3}). This can be
easily generalized to any pair of $s$-ordered \cite{Cahill69} two--time
correlations. Taking into account that the $s$-ordered two--time correlation
is nothing but%
\begin{equation}
\left\langle \alpha \left( t+\tau \right) ,\alpha \left( t\right)
\right\rangle _{s}\equiv \frac{1+s}{2}\left\langle \alpha \left( t+\tau
\right) ,\alpha \left( t\right) \right\rangle _{\mathcal{P}}+\frac{1-s}{2}%
\left\langle \alpha \left( t+\tau \right) ,\alpha \left( t\right)
\right\rangle _{Q},
\end{equation}%
with $s\in \left[ -1,1\right] $, it follows that%
\begin{equation}
\left\langle \alpha \left( t+\tau \right) ,\alpha \left( t\right)
\right\rangle _{\mathcal{P}}=\frac{1-s^{\prime }}{s-s^{\prime }}\left\langle
\alpha \left( t+\tau \right) ,\alpha \left( t\right) \right\rangle _{s}+%
\frac{1-s}{s^{\prime }-s}\left\langle \alpha \left( t+\tau \right) ,\alpha
\left( t\right) \right\rangle _{s^{\prime }}  \label{two-time gen}
\end{equation}%
for any $s\neq s^{\prime }$. Notice that Eq. (\ref{two-timeW}) is retrieved
from Eq. (\ref{two-time gen}) when $s=0$ (symmetric ordering, which is
obtained with the $W$ distribution) and $s^{\prime }=-1$ (antinormal
ordering, which is obtained with the $Q$ distribution). Now, following the
same arguments that lead to Eq. (\ref{Sout4}) one gets%
\begin{equation}
S_{\varphi }^{\mathrm{out}}\left( \omega \right) =\frac{1}{4}+\gamma _{%
\mathrm{out}}\left[ \frac{1-s^{\prime }}{s-s^{\prime }}\mathcal{V}_{s}\left(
\omega ,\varphi \right) +\frac{1-s}{s^{\prime }-s}\mathcal{V}_{s^{\prime
}}\left( \omega ,\varphi \right) \right] .  \label{Sout5}
\end{equation}%
We see that the use of the $\mathcal{P}$ distribution is equivalent to the
combined use of a pair of $s$-ordered distributions.

The interest of this approach is that Eqs. (\ref{Sout3}) and (\ref{Sout5})
provide a way for comparing the predictions of a pair of $s-$ordered
distributions, which we denote by $W_{s}$ ($W_{1}\equiv P$, $W_{0}\equiv W$, 
$W_{-1}\equiv Q$), with that of the $\mathcal{P}$ distribution. This is
interesting because the equation of evolution for a particular $W_{s}$ needs
not be of the Fokker--Planck type. For example, in the case we treat along
this article (dispersive optical bistability \cite%
{Drummond80b,Vogel88,Vogel89}), the equation of $W_{s}$ includes additional
terms (namely, third order derivatives) but for $s=\pm 1$, and, in general,
the diffusion matrix is not positive semidefinite \cite{Vogel88,Vogel89},
but for $s=0$. But these limitations do not necessarily prevent the use of
these distributions as under some reasonable approximations their equations
of evolution can be approximated to a Fokker--Planck equation (by neglecting
the higher order derivatives in the Wigner case \cite{Vogel88,Vogel89} or by
limiting the study to a parameter domain where the diffusion matrix is well
behaved in the Husimi case \cite{Savage88}). The point is that after making
these approximations, Eq. (\ref{Sout5}) should provide not an exact but an
approximate result. Then, by comparing the predictions of Eq. (\ref{Sout5})
to that of Eq. (\ref{Sout3}) one could evaluate the influence of these
approximations.

In this article we shall make use of these approximations for the special
case of dispersive optical bistability \cite{Drummond80b,Vogel88,Vogel89}.
We then derive the fluctuation spectra from the linearized Langevin
equations coming from $s-$ordered quasiprobability distributions, $W_{s}$,
and compute the (linearized) squeezing spectrum. The main result we obtain
is that, although any $W_{s}$ obeys an approximate Fokker--Planck equation,
and thus approximated Langevin equations can be obtained, the linearized
squeezing spectrum given by Eq.(\ref{Sout5}) is identical to that given by
Eq. (\ref{Sout3}). In other words, the approximations made in deriving
Langevin equations from the approximated equations of evolution do not
manifest in the linearized fluctuations spectra. We show further that the
predictions for the (linearized) intracavity squeezing from any $W_{s}$ is
also exact.

\section{Model for dispersive optical bistability}

\subsection{Master equation}

We shall adopt the model for dispersive optical bistability studied by
Drummond and Walls \cite{Drummond80b}, consisting of a single--ended optical
cavity containing a purely dispersive and isotropic $\chi ^{\left( 3\right)
} $ medium and pumped by a coherent field of frequency $\omega $ close to
that of a cavity mode, $\omega _{\mathrm{c}}$. The system Hamiltonian in the
interaction picture reads 
\begin{equation}
H=\hbar \left[ \left( \theta -g\right) \hat{a}^{\dagger }\hat{a}%
+iE_{0}\left( \hat{a}^{\dagger }-\hat{a}\right) -\frac{g}{2}\hat{a}^{\dagger
2}\hat{a}^{2}\right] ,  \label{Hamiltonian}
\end{equation}%
where $E_{0}$ is proportional to the amplitude of the injected field, $%
\theta =\omega _{\mathrm{c}}-\omega $ is a detuning, and $g\equiv
3\varepsilon _{0}\hbar \omega _{\mathrm{c}}^{2}\chi /\left( \varepsilon
^{2}V\right) $ is the coupling constant, with $V$ the quantization volume, $%
\varepsilon $ the medium dielectric constant and $\chi =\chi _{iiii}^{\left(
3\right) }$ ($i=1,2,3$) the nonlinear susceptibility \cite{Boyd}. We note
that we used a symmetrized Hamiltonian and this is the reason why the
detuning is not $\theta $ but $\left( \theta -g\right) $: Had it been
calculated in, say, normal or antinormal order, the detuning would have been 
$\theta $ and $\left( \theta -2g\right) $, respectively. Then, the
correction $g$ to the detuning is nothing but the modification of the cavity
frequency $\omega _{\mathrm{c}}$ due to vacuum fluctuations as described
with the different ordering choices.

The intracavity field mode exits the cavity through the output mirror.
Assuming weak coupling between the field mode and the rest of vacuum modes,
which are treated as a reservoir (see, e.g., \cite{Carmichael99}), the
master equation of the system at zero temperature reads \cite{Drummond80b}%
\begin{align}
\dot{\rho}& =E_{0}\left( a^{\dagger }\rho -\rho a^{\dagger }+\rho a-a\rho
\right) -i\left( \theta -g\right) \left( a^{\dagger }a\rho -\rho a^{\dagger
}a\right) \\
& +i\frac{g}{2}\left( a^{\dagger 2}a^{2}\rho -\rho a^{\dagger 2}a^{2}\right)
+\frac{\gamma }{2}\left( 2a\rho a^{\dagger }-\rho a^{\dagger }a-a^{\dagger
}a\rho \right) ,  \notag
\end{align}%
where $\gamma $ represents the cavity losses of the field intensity and, as
the cavity is single-ended, $\gamma _{\mathrm{out}}=\gamma $.

\subsection{Quasiprobability distributions}

The equation of evolution for the $s$-ordered quasiprobability distribution
for dispersive optical bistability was first derived by Vogel and Risken 
\cite{Vogel89}. With our notation%
\begin{align}
\frac{\partial }{\partial t}W_{s}\left( \alpha ,\alpha ^{\ast }\right) & =%
\left[ -\sum_{i}\frac{\partial }{\partial \alpha _{i}}A_{i}^{\left( s\right)
}+\frac{1}{2}\sum_{i,j}\frac{\partial ^{2}}{\partial \alpha _{i}\partial
\alpha _{j}}D_{ij}^{\left( s\right) }\right] W_{s}  \label{evol total} \\
& +ig\frac{1-s^{2}}{4}\left( \frac{\partial ^{3}}{\partial \alpha
^{2}\partial \alpha ^{\ast }}\alpha -\frac{\partial ^{3}}{\partial \alpha
\partial \alpha ^{\ast 2}}\alpha ^{\ast }\right) W_{s},  \notag
\end{align}%
where $\alpha _{1}\equiv \alpha $, $\alpha _{2}\equiv \alpha ^{\ast }$, and%
\begin{align}
A_{1}^{\left( s\right) }& \equiv E_{0}-\left[ \frac{\gamma }{2}+i\left(
\theta -sg\right) \right] \alpha +ig\alpha ^{2}\alpha ^{\ast },  \label{A1s}
\\
A_{2}^{\left( s\right) }& \equiv E_{0}-\left[ \frac{\gamma }{2}-i\left(
\theta -sg\right) \right] \alpha ^{\ast }-ig\alpha \left( \alpha ^{\ast
}\right) ^{2},  \label{A2s} \\
\mathbb{D}^{\left( s\right) }& \equiv \left( 
\begin{array}{cc}
isg\alpha ^{2} & \frac{1-s}{2}\gamma \\ 
\frac{1-s}{2}\gamma & -isg\left( \alpha ^{\ast }\right) ^{2}%
\end{array}%
\right) .  \label{Dgen}
\end{align}%
($\left[ \mathbb{D}^{\left( s\right) }\right] _{ij}\equiv D_{ij}^{\left(
s\right) }.$) Making use of Eq. (\ref{Dgen}) we obtain%
\begin{equation}
\mathbb{D}^{\left( 1\right) }=\frac{1-s^{\prime }}{s-s^{\prime }}\mathbb{D}%
^{\left( s\right) }+\frac{1-s}{s^{\prime }-s}\mathbb{D}^{\left( s^{\prime
}\right) }.  \label{propiedad-gen}
\end{equation}%
We shall make use of this property later.

For $s=1,0,-1$, $W_{s}$ corresponds to the $P$ (Glauber--Sudarshan), $W$
(Wigner), and $Q$ (Husimi) distributions respectively as commented. An
alternative quasiprobability distribution to $W_{s}$ is the so--called
generalized $P$ distribution \cite{Drummond80}, which we have already
denoted by $\mathcal{P}$. Its equation of evolution has been derived by
Drummond and Walls \cite{Drummond80b}. It is given by Eq. (\ref{evol total})
with $s=1$ after changing $\alpha ^{\ast }$ by the complex variable $\beta $%
, which is independent of $\alpha $ and verifies $\left\langle \alpha
\right\rangle =\left\langle \beta \right\rangle ^{\ast }$ \cite{Drummond80}.
Hence in this representation the phase space is doubled with respect to the $%
W_{s}$ representation.

\subsection{Fokker-Planck equation for the $W_{s}$ distribution}

By construction all three $\mathcal{P}$, $P$, and $Q$ quasiprobability
distributions formally obey a Fokker--Planck equation \cite%
{Walls94,Carmichael99} as they do not contain derivatives of order higher
than $2$. This is not the case for any $W_{s}$ with $s\neq \pm 1$. We show
next that, in spite of this fact, any $W_{s}$ distribution verifies an
approximate Fokker--Planck equation. This statement is equivalent to saying
that, in some limit, the third order derivatives in Eq.(\ref{evol total}),
existing unless $s=\pm 1$, can be neglected. For that we make a system size
expansion \cite{Carmichael99,Vogel89}, which is based on the very large
value attained by the mean number of intracavity photons $\left\langle
a^{\dagger }a\right\rangle \sim \left\vert \alpha \right\vert ^{2}\sim
\gamma /\left\vert g\right\vert $ \cite{Vogel88,Vogel89}. For example, by
taking $V=1\mathrm{cm}^{3}$, $\left\vert \chi \right\vert =5\cdot 10^{-23}%
\mathrm{m}^{2}\mathrm{Volt}^{-2}$, $\varepsilon =4\varepsilon _{0}$, $\omega
_{\mathrm{c}}=3\cdot 10^{15}\mathrm{s}^{-1}$, one obtains $\left\vert
g\right\vert \sim 10^{-9}\mathrm{s}^{-1}$ and, taking $\gamma =10^{9}\mathrm{%
s}^{-1}$, one has $\gamma /\left\vert g\right\vert \sim 10^{18}$. Now,
normalizing time to $\gamma $ and $\alpha $ to $\sqrt{\gamma /g}$, one
obtains an equation equivalent to Eq. (\ref{evol total}) in which the third
order derivatives are multiplied by $\left( g/\gamma \right) ^{2}$, whilst
the second order derivatives are multiplied by $g/\gamma $ and the first
order derivatives are of order one. Then, the neglection of the third order
derivatives looks like a very accurate approximation. Notice, however, that
the predictions of such a truncated equation may differ significantly from
their correct values, as it occurs with the tunneling times \cite%
{Vogel88,Vogel89}. Finally note that, given the smallest value of $g$, all $%
A_{i}^{\left( s\right) }$ [Eqs. (\ref{A1s},\ref{A2s})] can be approximated
by $A_{i}^{\left( s=0\right) }$ \cite{Vogel88,Vogel89}, which we denote just
by $A_{i}$. Once the system size expansion has been performed Eq. (\ref{evol
total}) becomes%
\begin{eqnarray}
\frac{\partial }{\partial t}W_{s} &\simeq &\left[ -\sum_{i}\frac{\partial }{%
\partial \alpha _{i}}A_{i}+\frac{1}{2}\sum_{i,j}\frac{\partial ^{2}}{%
\partial \alpha _{i}\partial \alpha _{j}}D_{ij}^{\left( s\right) }\right]
W_{s},  \label{FPs} \\
A_{1} &\equiv &E_{0}-\left( \frac{\gamma }{2}+i\theta \right) \alpha
+ig\alpha ^{2}\alpha ^{\ast },  \label{A1} \\
A_{2} &\equiv &E_{0}-\left( \frac{\gamma }{2}-i\theta \right) \alpha ^{\ast
}-ig\alpha \left( \alpha ^{\ast }\right) ^{2},  \label{A2}
\end{eqnarray}%
$\alpha _{1}\equiv \alpha $, $\alpha _{2}\equiv \alpha ^{\ast }$, and $%
\mathbb{D}^{\left( s\right) }$ is given by Eq. (\ref{Dgen}). The symbol $%
\simeq $ is used instead of the equality symbol in order to stress that Eq. (%
\ref{FPs}) is approximate (but for $s=\pm 1$).

As announced Eq. (\ref{FPs}) is a Fokker--Planck equation. However it is in
fact a pseudo Fokker--Plank equation \cite{Vogel88,Vogel89} as the diffusion
matrix $\mathbb{D}^{\left( s\right) }$ is not positive semidefinite in
general and then the equation cannot be interpreted as describing a
generalized Brownian motion. That $\mathbb{D}^{\left( s\right) }$ is not
positive semidefinite is easy to see by writing Eq. (\ref{FPs}) in terms of
the real variables $x=\func{Re}\alpha $, $y=\func{Im}\alpha $. For these new
variables an equation similar to Eq. (\ref{FPs}) is obtained with a
diffusion matrix $\mathbb{D}_{xy}^{\left( s\right) }$ given by \cite{Vogel88}%
\begin{equation*}
\mathbb{D}_{xy}^{\left( s\right) }\equiv \left( 
\begin{array}{cc}
\frac{1-s}{4}\gamma -sgxy & \frac{sg}{2}\left( x^{2}-y^{2}\right) \\ 
\frac{sg}{2}\left( x^{2}-y^{2}\right) & \frac{1-s}{4}\gamma +sgxy%
\end{array}%
\right) ,
\end{equation*}%
whose eigenvalues $d_{\pm }^{\left( s\right) }$ read%
\begin{equation*}
d_{\pm }^{\left( s\right) }=\frac{1-s}{4}\gamma \pm \frac{\left\vert
sg\right\vert }{2}\left( x^{2}+y^{2}\right) .
\end{equation*}%
The positive semidefiniteness of $\mathbb{D}^{\left( s\right) }$ then
requires that $d_{-}^{\left( s\right) }\geq 0$:%
\begin{equation}
\left\vert \alpha \right\vert ^{2}\leq \frac{\gamma }{\left\vert
g\right\vert }\frac{1-s}{2\left\vert s\right\vert }.  \label{semipos}
\end{equation}%
(Remind that $\left\vert \alpha \right\vert ^{2}=x^{2}+y^{2}$.) Clearly,
only for $s=0$ (Wigner distribution) condition (\ref{semipos}) is fulfilled
for any $\alpha $. On the other hand for $s=+1$ ($P$ distribution) condition
(\ref{semipos}) is never satisfied. In general, for any $s\neq 0,1$ that
condition is verified inside a bounded region of the phase space and thus, $%
\mathbb{D}^{\left( s\right) }$ is never, strictly speaking, positive
semidefinite. Nevertheless if $\alpha $ is replaced by its classical steady
value $\bar{\alpha}$, what is done for calculating linearized spectra as we
do here (see below), $\mathbb{D}^{\left( s\right) }$ will be positive
semidefinite whenever condition (\ref{semipos}) holds when applied to the
classical steady state. This restricted condition is in fact verified in a
bounded region of the parameter space. We note that this approximation was
done for the case of second--harmonic generation by Savage \cite{Savage88}.
It can be understood in the sense that one assumes that $W_{s}$ is peaked
around the steady state value and that the parameters of the system are such
that a negligible part of the distribution violates the condition $%
d_{-}^{\left( s\right) }\geq 0$. Under this approximation $\mathbb{D}%
^{\left( s\right) }$ is a well behaved diffusion matrix and the equation of
evolution of $W_{s}$ is a true Fokker--Planck equation for any $s\left( \neq
+1\right) $.

The requirement of positive semidefiniteness of the diffusion matrix $%
\mathbb{D}^{\left( s\right) }$ comes from the fact that $\alpha $ and $%
\alpha ^{\ast }$ are complex--conjugate variables, and the noise terms in
the Langevin equations equivalent to the Fokker--Planck equation (see below)
will not be complex--conjugated if this requirement is not fulfilled. Notice
that this requirement is lifted in the case of the $\mathcal{P}$
distribution as $\alpha $ and $\beta $ are not complex--conjugate variables,
but in their mean.

\section{Linearized fluctuations spectra and squeezing}

The Ito stochastic differential equations that are equivalent to the
Fokker--Planck equation (\ref{FPs}), assuming the positive semidefiniteness
of the diffusion matrix, are%
\begin{equation}
\frac{d\mathbf{\alpha }}{dt}\simeq \mathbf{A}\left( \mathbf{\alpha }\right) +%
\mathbb{B}^{\left( s\right) }\left( \mathbf{\alpha }\right) \cdot \mathbf{%
\xi }\left( t\right) ,  \label{nonlinearL}
\end{equation}%
where $\mathbf{\alpha }\equiv \left( \alpha ,\alpha ^{\ast }\right) ^{%
\mathrm{T}}$, $\mathbf{A}\equiv \left( A_{1},A_{2}\right) ^{\mathrm{T}}$, $%
\mathbb{D}^{\left( s\right) }\left( \mathbf{\alpha }\right) \equiv \mathbb{B}%
^{\left( s\right) }\left( \mathbf{\alpha }\right) \left[ \mathbb{B}^{\left(
s\right) }\left( \mathbf{\alpha }\right) \right] ^{\mathrm{T}}$, and the
white Gaussian noise term $\mathbf{\xi }\equiv \left( \xi _{1},\xi
_{2}\right) ^{\mathrm{T}}$ verifies $\left\langle \xi _{i}\left( t\right)
\right\rangle =0$, and $\left\langle \xi _{i}\left( t\right) \xi _{j}\left(
t^{\prime }\right) \right\rangle =\delta _{ij}\delta \left( t-t^{\prime
}\right) $. (We note that $\mathbf{\alpha }$ should contain the label $s$,
e.g. $\mathbf{\alpha }^{\left( s\right) }$, as that stochastic variable is
representation dependent. We avoid this labeling in order to not overburden
the notation.)

In this article we shall limit ourselves to the study of fluctuations around
the classical steady state $\mathbf{\alpha }=\mathbf{\bar{\alpha}}$ (the
solution to $\mathbf{A}\left( \mathbf{\bar{\alpha}}\right) =0$) in the
linear approximation. The linearized Langevin equations read%
\begin{align}
\frac{d}{dt}\delta \mathbf{\alpha }& \simeq \mathbb{\bar{A}}\cdot \delta 
\mathbf{\alpha }+\mathbb{\bar{B}}^{\left( s\right) }\cdot \mathbf{\xi }%
\left( t\right) ,  \label{linearL} \\
\left[ \mathbb{\bar{A}}\right] _{ij}& \equiv \left( \frac{\partial A_{i}}{%
\partial \alpha _{j}}\right) _{\mathbf{\alpha }=\mathbf{\bar{\alpha}}},~\ 
\mathbb{\bar{B}}^{\left( s\right) }\equiv \mathbb{B}^{\left( s\right)
}\left( \mathbf{\alpha }=\mathbf{\bar{\alpha}}\right) ,
\end{align}%
$\alpha _{1}\equiv \alpha $ and $\alpha _{2}\equiv \alpha ^{\ast }$.

We are concerned with the calculation of the spectral matrix of
fluctuations, $\mathbb{S}^{\left( s\right) }\left( \omega \right) $, of
elements%
\begin{equation}
S_{ij}^{\left( s\right) }\left( \omega \right) \equiv \int_{-\infty
}^{+\infty }d\tau e^{-i\omega \tau }\left\langle \alpha _{i}\left( t+\tau
\right) ,\alpha _{j}\left( t\right) \right\rangle _{s},  \label{Sij}
\end{equation}%
(not to be confused with the squeezing spectrum) which, from the linearized
Langevin equations (\ref{linearL}), can be easily obtained by making use of 
\cite{Chaturvedi77}%
\begin{eqnarray}
\mathbb{S}^{\left( s\right) }\left( \omega \right) &\simeq &\mathbb{S}_{%
\mathrm{approx}}^{\left( s\right) }\left( \omega \right) ,  \label{Sapprox}
\\
\mathbb{S}_{\mathrm{approx}}^{\left( s\right) }\left( \omega \right) &\equiv
&\left( \mathbb{\bar{A}}+i\mathbb{\omega I}\right) ^{-1}\mathbb{\bar{D}}%
^{\left( s\right) }\left( \mathbb{\bar{A}}^{\mathrm{T}}-i\mathbb{\omega I}%
\right) ^{-1},  \label{chaturvedi}
\end{eqnarray}%
where $\mathbb{I}$ denotes the $2\times 2$ identity matrix and $\mathbb{\bar{%
D}}^{\left( s\right) }=\mathbb{D}^{\left( s\right) }\left( \mathbf{\alpha }=%
\mathbf{\bar{\alpha}}\right) $ and again the symbol $\simeq $ stresses that
the result is an approximation. The result for the different $s-$orderings
is given in Appendix A.

We remind that the Fokker--Planck equation obeyed by $\mathcal{P}$ is given
by Eq. (\ref{FPs}) for $s=1$ with the replacement $\alpha _{2}=\alpha ^{\ast
}\rightarrow \beta $. Then the full and linearized Langevin equations in the 
$\mathcal{P}$ representation are given by Eqs. (\ref{nonlinearL}) and (\ref%
{linearL}) respectively, under the previous replacement, and the spectral
matrix corresponding to the $\mathcal{P}$ distribution, $\mathbb{S}^{%
\mathcal{P}}$, is given by Eq. (\ref{chaturvedi}) with $s=1$ (note that $%
\bar{\beta}=\bar{\alpha}^{\ast }$):%
\begin{equation}
\mathbb{S}^{\mathcal{P}}\left( \omega \right) =\left( \mathbb{\bar{A}}+i%
\mathbb{\omega I}\right) ^{-1}\mathbb{\bar{D}}^{\left( 1\right) }\left( 
\mathbb{\bar{A}}^{\mathrm{T}}-i\mathbb{\omega I}\right) ^{-1}.
\label{chaturvediP}
\end{equation}%
We also remind that Eq. (\ref{FPs}) for $s=1$ is exact as the original Eq. (%
\ref{evol total}) does not contain third order derivatives in this case;
hence all symbols $\simeq $ must be replaced by $=$ in this case, as in Eq. (%
\ref{chaturvediP}). On the other hand the normally ordered spectral matrix
of fluctuations defined as%
\begin{equation*}
:\mathbb{S}\left( \omega \right) :~\equiv ~:\int_{-\infty }^{+\infty }d\tau
e^{-i\omega \tau }%
\begin{pmatrix}
\left\langle \hat{a}\left( t+\tau \right) ,\hat{a}\left( t\right)
\right\rangle & \left\langle \hat{a}\left( t+\tau \right) ,\hat{a}^{\dagger
}\left( t\right) \right\rangle \\ 
\left\langle \hat{a}^{\dagger }\left( t+\tau \right) ,\hat{a}\left( t\right)
\right\rangle & \left\langle \hat{a}^{\dagger }\left( t+\tau \right) ,\hat{a}%
^{\dagger }\left( t\right) \right\rangle%
\end{pmatrix}%
:
\end{equation*}%
equals, by definition, $\mathbb{S}^{\mathcal{P}}\left( \omega \right) $.
Thus Eq. (\ref{chaturvediP}) yields the exact normally ordered spectral
matrix of fluctuations,%
\begin{equation}
:\mathbb{S}\left( \omega \right) :~=\mathbb{S}^{\mathcal{P}}\left( \omega
\right) .  \label{S=S}
\end{equation}%
We now recall property (\ref{two-time gen}) that, together with definition (%
\ref{Sij}), allows to state that%
\begin{equation}
\mathbb{S}^{\mathcal{P}}\left( \omega \right) =\frac{1-s^{\prime }}{%
s-s^{\prime }}\mathbb{S}^{\left( s\right) }\left( \omega \right) +\frac{1-s}{%
s^{\prime }-s}\mathbb{S}^{\left( s^{\prime }\right) }\left( \omega \right) ,
\label{P=SSprime}
\end{equation}%
which, making use of Eq. (\ref{Sapprox}), can be approximated as%
\begin{equation}
\mathbb{S}^{\mathcal{P}}\left( \omega \right) \simeq \frac{1-s^{\prime }}{%
s-s^{\prime }}\mathbb{S}_{\mathrm{approx}}^{\left( s\right) }\left( \omega
\right) +\frac{1-s}{s^{\prime }-s}\mathbb{S}_{\mathrm{approx}}^{\left(
s^{\prime }\right) }\left( \omega \right) .  \label{P=SSprimeapprox}
\end{equation}%
Now, substituting Eq. (\ref{chaturvedi}) into (\ref{P=SSprimeapprox}), and
recalling property (\ref{propiedad-gen}) and Eq. (\ref{chaturvediP}), we
observe that the approximate equality (\ref{P=SSprimeapprox}) is a true
equality indeed.

This means that, although the used Langevin equations come from approximated
(truncated in general) Fokker--Planck equations, the spectral matrices $%
\mathbb{S}_{\mathrm{approx}}^{\left( s\right) }\left( \omega \right) $ given
by Eq. (\ref{chaturvedi}) provide the correct result. In other words, the
third order derivatives present in the original pseudo Fokker--Planck
equation (\ref{evol total}) seem to play no role on the correlations between
fluctuations in a linearized theory. Moreover, the approximated Langevin
equations (\ref{nonlinearL}) ignore that the diffusion matrices $\mathbb{D}%
^{\left( s\right) }$ can be non positive semidefinite, as discussed above.
Nevertheless, as relation (\ref{P=SSprimeapprox}) is not approximate but
exact, and it holds for any parameter set, even where $\mathbb{D}^{\left(
s\right) }$ is not positive semidefinite, we conclude that the positive
semidefinite condition on $\mathbb{D}^{\left( s\right) }$ is irrelevant in
the following sense. We recall that the same occurs with the
Glauber--Sudarshan $P$ distribution: even if the pseudo Fokker--Planck
equation governing its evolution has a non positive semidefinite diffusion
matrix $\mathbb{D}^{\left( 1\right) }$, one can nevertheless write down a
corresponding Langevin equation, which yields the correct result for the
spectral matrix $\mathbb{S}$ \cite{Drummond80}. The explanation for this was
given by Drummond by introducing the generalized $P$ representation \cite%
{Drummond80}, which operationally amounts to substitute the complex
conjugated variable $\alpha ^{\ast }$ in the pseudo Fokker--Planck equation
verified by the Glauber--Sudarshan $P$ by an independent complex variable $%
\beta $. This suggests that our result can be understood in terms of
"generalized $W_{s}$ distributions", call them $\mathcal{W}_{s}$: Should we
substitute $\alpha ^{\ast }$ by an independent complex variable $\beta $ in
the original pseudo Fokker--Planck equation (\ref{evol total}) a positive
semidefinite diffusion matrix would be not needed in order to derive
corresponding Langevin equations (once the third order derivatives had been
neglected). These Langevin equations for $\mathcal{W}_{s}$ would read as
those for $W_{s}$ but with $\alpha ^{\ast }\rightarrow \beta $, and the
final expression for the spectral matrix, which would be exact in this
generalized representation, would be given by Eq. (\ref{chaturvedi}) in the
linear approximation, just as it happens in our case. The possibility of
defining "generalized $W_{s}$ distributions" should be studied, probably by
defining $\mathcal{W}_{s}$ in terms of $\mathcal{P}$, as $W_{s}$ is defined
in terms of $P$ \cite{Cahill69}. We leave this discussion open as it is out
of the scope of the present work.

The above discussion implies that the squeezing spectra of the output field,
given by Eqs. (\ref{Sout3}) and (\ref{Sout5}) with%
\begin{eqnarray}
\mathcal{V}_{\mathcal{P}}\left( \omega ,\varphi \right) &=&\frac{1}{4}\left[
S_{11}^{\mathcal{P}}\left( \omega \right) e^{-2i\varphi }+S_{22}^{\mathcal{P}%
}\left( \omega \right) e^{+2i\varphi }+S_{12}^{\mathcal{P}}\left( \omega
\right) +S_{21}^{\mathcal{P}}\left( \omega \right) \right] , \\
\mathcal{V}_{s}\left( \omega ,\varphi \right) &=&\frac{1}{4}\left[
S_{11}^{(s)}\left( \omega \right) e^{-2i\varphi }+S_{22}^{(s)}\left( \omega
\right) e^{+2i\varphi }+S_{12}^{(s)}\left( \omega \right)
+S_{21}^{(s)}\left( \omega \right) \right] ,  \label{Vs}
\end{eqnarray}%
are, obviously, the same. We shall not analyze here the properties of this
squeezing spectrum as this analysis can be found in \cite{Walls94}. We just
quote in Appendix B the expression for $\mathcal{V}_{s}\left( \omega
,\varphi \right) $. Nevertheless we want to make a comment on the amount of
squeezing attainable inside the nonlinear cavity. This can be calculated by
integrating the spectrum of fluctuations of the field quadratures, i.e.%
\begin{equation}
V_{s}\left( \varphi \right) =\frac{1}{2\pi }\int_{-\infty }^{+\infty
}d\omega \mathcal{V}_{s}\left( \omega ,\varphi \right) .
\end{equation}%
We must take into account how the different $V_{s}$ are related with the
squeezing $V\left( \varphi \right) \equiv \left\langle \hat{X}_{\varphi
}\left( t\right) ,\hat{X}_{\varphi }\left( t\right) \right\rangle $ where $%
\hat{X}_{\varphi }$ is defined as $\hat{X}_{\varphi }^{\mathrm{out}}$, Eq. (%
\ref{quadratures}), but replacing $\left( \hat{a}_{\mathrm{out}},\hat{a}_{%
\mathrm{out}}^{\dagger }\right) $ with $\left( \hat{a},\hat{a}^{\dagger
}\right) $. Performing the calculation one easily obtains, with the help of
the commutator $\left[ \hat{a}\left( t\right) ,\hat{a}^{\dagger }\left(
t\right) \right] =1$ the result%
\begin{equation}
V\left( \varphi \right) =V_{s}\left( \varphi \right) +\frac{s}{4}.
\end{equation}%
The point is that, as any $\mathcal{V}_{s}\left( \omega ,\varphi \right) $,
Eq. (\ref{Vs}), can be computed from $\mathbb{S}_{\mathrm{approx}}^{\left(
s\right) }\left( \omega \right) $, Eq. (\ref{Sapprox}), and $\mathbb{S}_{%
\mathrm{approx}}^{\left( s\right) }\left( \omega \right) $ yields the
correct result, any single $W_{s}$ is useful for computing the intracavity
squeezing. We note that $V\left( \varphi \right) $ is minimum for a
particular $\varphi $ and is $V_{\min }=\frac{1}{8}$. This is a well known
result: The maximum degree of squeezing attainable inside a nonlinear
cavity, which happens at the bifurcation points, is half that of a coherent
state.

\section{Conclusions}

In this article we have discussed how the spectrum of squeezing of the field
outgoing a nonlinear cavity can be derived from a combination of the spectra
of intracavity fluctuations obtained from Langevin equations derived from $%
W_{s}$ distributions. We have illustrated this for the special case of
dispersive optical bistability. The interesting result is that the
linearized spectrum of squeezing obtained is this way is exact in spite of
the fact that no $W_{s}$ quasiprobability distribution verifies
Fokker--Planck equations but only approximate ones. We have also shown that
the predictions for the (linearized) squeezing attainable inside the
nonlinear cavity is correct when calculated with any $W_{s}$. The conclusion
is that the linearized Langevin equations corresponding to a $W_{s}$
representation are correct or, in other words, that the approximations made
for converting the equation of evolution of $W_{s}$ into Fokker--Planck
equations do not manifest in the linearized theory, even if the diffusion
matrix of the Fokker--Planck equation is not positive semidefinite. The
latter has allowed us to conjecture the definition of "generalized $W_{s}$
distributions", following the spirit of the generalized $P$ distributions of
Drummond. Of course, when going to the nonlinear regime, as in the
calculation of e.g. tunneling times, one must be cautious about the
truncation of pseudo Fokker--Planck equations containing third (or higher
order) derivatives.

\section{Acknowledgements}

This work was supported by the Spanish Ministerio de Ciencia y Tecnolog\'{\i}%
a and the European Union FEDER (Project No. BFM2002-04369-C04-01).

\section{Appendix A}

In this Appendix the expression for the spectral matrix $\mathbb{S}_{\mathrm{%
approx}}^{\left( s\right) }\left( \omega \right) $ defined in Eq. (\ref%
{chaturvedi}) is given for any $s-$ordered quasidistribution $W_{s}$. The
calculation needs the computation of the classical steady state $\left(
\alpha =\bar{\alpha},\alpha ^{\ast }=\bar{\alpha}^{\ast }\right) $, which is
given by $A_{1}=A_{2}=0$, Eqs. (\ref{A1}) and (\ref{A2}). After introducing
the quantities 
\begin{equation}
\Delta =\frac{2\eta \theta }{\gamma },~\mu =\left( \frac{2}{\gamma }\right)
^{3}\left\vert g\right\vert E_{0}^{2},~\sqrt{I}e^{i\phi }=\left( \frac{%
2\left\vert g\right\vert }{\gamma }\right) ^{\frac{1}{2}}\bar{\alpha},~\eta =%
\func{sign}\left( g\right) ,  \label{cambios}
\end{equation}%
the classical steady state is given by%
\begin{eqnarray}
\mu &=&I\left[ 1+\left( I-\Delta \right) ^{2}\right] ,  \label{muI} \\
e^{i\phi } &=&\frac{1+i\eta \left( I-\Delta \right) }{\sqrt{1+\left(
I-\Delta \right) ^{2}}}.
\end{eqnarray}%
The characteristic $I$ vs. $\mu $ displays bistable behaviour for $\Delta >%
\sqrt{3}$ as is well known, and the values of the intensity $I$ at the
turning points of the characteristic, $I=I_{\pm }$, are given by%
\begin{equation}
I_{\pm }\equiv \frac{2\Delta \pm \sqrt{\Delta ^{2}-3}}{3}.  \label{turning}
\end{equation}%
For $I_{-}<I<I_{+}$ the steady state is unstable; otherwise it is linearly
stable. As the state equation (\ref{muI}) implies that the pump power $\mu $
is univocally determined by $I$ one can use the latter as the control
parameter, and this is more convenient mathematically.

Making use of Eq. (\ref{chaturvedi}) one readily obtains $\left( \Omega
\equiv 2\omega /\gamma \right) $:%
\begin{align}
S_{11}^{\left( s\right) }\left( \omega \right) & =\frac{2}{\gamma }%
Ie^{2i\phi }\frac{2\left( \Delta -2I\right) +i\eta \left[ 2+s\left( \Omega
^{2}-\mathcal{I}\right) \right] }{\left( \Omega ^{2}-\mathcal{I}\right)
^{2}+4\Omega ^{2}},  \label{S11} \\
S_{12}^{\left( s\right) }\left( \omega \right) & =\frac{2}{\gamma }\frac{%
2I^{2}+\left( 1-s\right) \left[ \Omega ^{2}+\mathcal{I}-2\eta \Omega \left(
2I-\Delta \right) \right] }{\left( \Omega ^{2}-\mathcal{I}\right)
^{2}+4\Omega ^{2}},  \label{S12} \\
S_{22}^{\left( s\right) }\left( \omega \right) & =\left[ S_{11}^{\left(
s\right) }\left( \omega \right) \right] ^{\ast },~S_{21}^{\left( s\right)
}\left( \omega \right) =S_{12}^{\left( s\right) }\left( -\omega \right) ,
\label{S22S21}
\end{align}%
where%
\begin{equation}
\mathcal{I}=3(I-I_{+})(I-I_{-}).  \label{zeta}
\end{equation}%
Note that at the turning points of the characteristic $(I=I_{+}$ or $%
I=I_{-}) $ $\mathcal{I}=0$ and all $S_{ij}^{\left( s\right) }\left( \omega
\right) $ diverge at $\omega =0$.

\section{Appendix B}

The expression for $\mathcal{V}_{s}\left( \omega ,\varphi \right) $, Eq. (%
\ref{Vs}), making use of Eqs. (\ref{S11})--(\ref{zeta}) in Appendix A, reads

\begin{equation}
\gamma \mathcal{V}_{s}\left( \omega ,\varphi \right) =\frac{2I\left( \Delta
-2I\right) \cos \psi -\eta I\left[ 2+s\left( \Omega ^{2}-\mathcal{I}\right) %
\right] \sin \psi +2I^{2}+\left( 1-s\right) \left( \Omega ^{2}+\mathcal{I}%
\right) }{\left( \Omega ^{2}-\mathcal{I}\right) ^{2}+4\Omega ^{2}}
\end{equation}%
where $\psi \equiv 2\left( \phi -\varphi \right) $.

As discussed $\mathcal{V}_{s}\left( \omega ,\varphi \right) $ for $s=+1$
coincides with the corresponding expression calculated in the $\mathcal{P}$
representation:%
\begin{equation}
\gamma \mathcal{V}_{\mathcal{P}}\left( \omega ,\varphi \right) =\frac{%
2I\left( \Delta -2I\right) \cos \psi -\eta I\left[ 2+\left( \Omega ^{2}-%
\mathcal{I}\right) \right] \sin \psi +2I^{2}}{\left( \Omega ^{2}-\mathcal{I}%
\right) ^{2}+4\Omega ^{2}}.
\end{equation}%
Note that this quantity is just $S_{\varphi }^{\mathrm{out}}\left( \omega
\right) -\frac{1}{4}$, Eq. (\ref{Sout3}) (remind that $\gamma _{\mathrm{out}%
}=\gamma $).

\end{document}